\documentclass[10pt,twocolumn,letterpaper]{article}

\usepackage{cvpr}              

\usepackage{graphicx}
\usepackage{amsmath}
\usepackage{amssymb}
\usepackage{booktabs}
\usepackage{graphicx}
\usepackage{multirow}
\usepackage{subcaption}
\usepackage{algorithm}
\usepackage{algpseudocode} 
\usepackage{threeparttable}
\usepackage[accsupp]{axessibility}  

\newcommand{\bheading}[1]{{\noindent{\textbf{#1}}}}

\newcommand{\tabincell}[2]{\begin{tabular}{@{}#1@{}}#2\end{tabular}}

\usepackage[pagebackref,breaklinks,colorlinks]{hyperref}

\usepackage[capitalize]{cleveref}
\crefname{section}{Sec.}{Secs.}
\Crefname{section}{Section}{Sections}
\Crefname{table}{Table}{Tables}
\crefname{table}{Tab.}{Tabs.}

\def\cvprPaperID{04486} 
\def\confName{CVPR}
\def\confYear{2022}

\begin{document}

\title{Which images to label for few-shot medical landmark detection?}
%

\author{Quan Quan\textsuperscript{1,3} \quad
	Qingsong Yao\textsuperscript{1,3}\quad
	Jun Li\textsuperscript{1,3}\quad
	S.Kevin Zhou$^{1,2}$ \\
	$^1$ Key Lab of Intelligent Information Processing of Chinese Academy \\ of Sciences (CAS), Institute of Computing Technology \\
	$^2$ University of Science and Technology of China \\
    $^3$ University of Chinese Academy of  Sciences \\
}

\maketitle

\begin{abstract}
	The success of deep learning methods relies on the availability of well-labeled large-scale datasets.
	However, for medical images, annotating such abundant training data often requires experienced radiologists and consumes their limited time.
	Few-shot learning is developed to alleviate this burden, which achieves competitive performances with only several labeled data. However, a crucial yet \textbf{previously overlooked} problem in few-shot learning is about the selection of template images for annotation before learning, which affects the final performance. 
	We herein propose a novel Sample Choosing Policy (SCP) to select “the most worthy" images for annotation, in the context of few-shot medical landmark detection.
	SCP consists of three parts: 1) Self-supervised training for building a pre-trained deep model to extract features from radiological images, 2) Key Point Proposal for localizing informative patches, and 3) Representative Score Estimation for searching the most representative samples or templates. 
	The advantage of SCP is demonstrated by various experiments on three widely-used public datasets. For one-shot medical landmark detection, its use reduces the mean radial errors on Cephalometric and HandXray datasets by 14.2\% (from 3.595mm to 3.083mm) and 35.5\% (4.114mm to 2.653mm), respectively. 
	
\end{abstract}

\section{Introduction}
\noindent 

It is widely known that the success of deep learning relies on data availability. Learning from datasets of a larger quantity and higher quality likely brings a higher performance and better generalization for neural networks. Yet, the labeling of datasets needs well-trained, highly-engaged radiologists for medical image analysis tasks~\cite{zhou2021review,zhou2019handbook}, which is especially challenging as physicians are costly and busy. 

\begin{figure}
	\centering
	\includegraphics[width=1\linewidth]{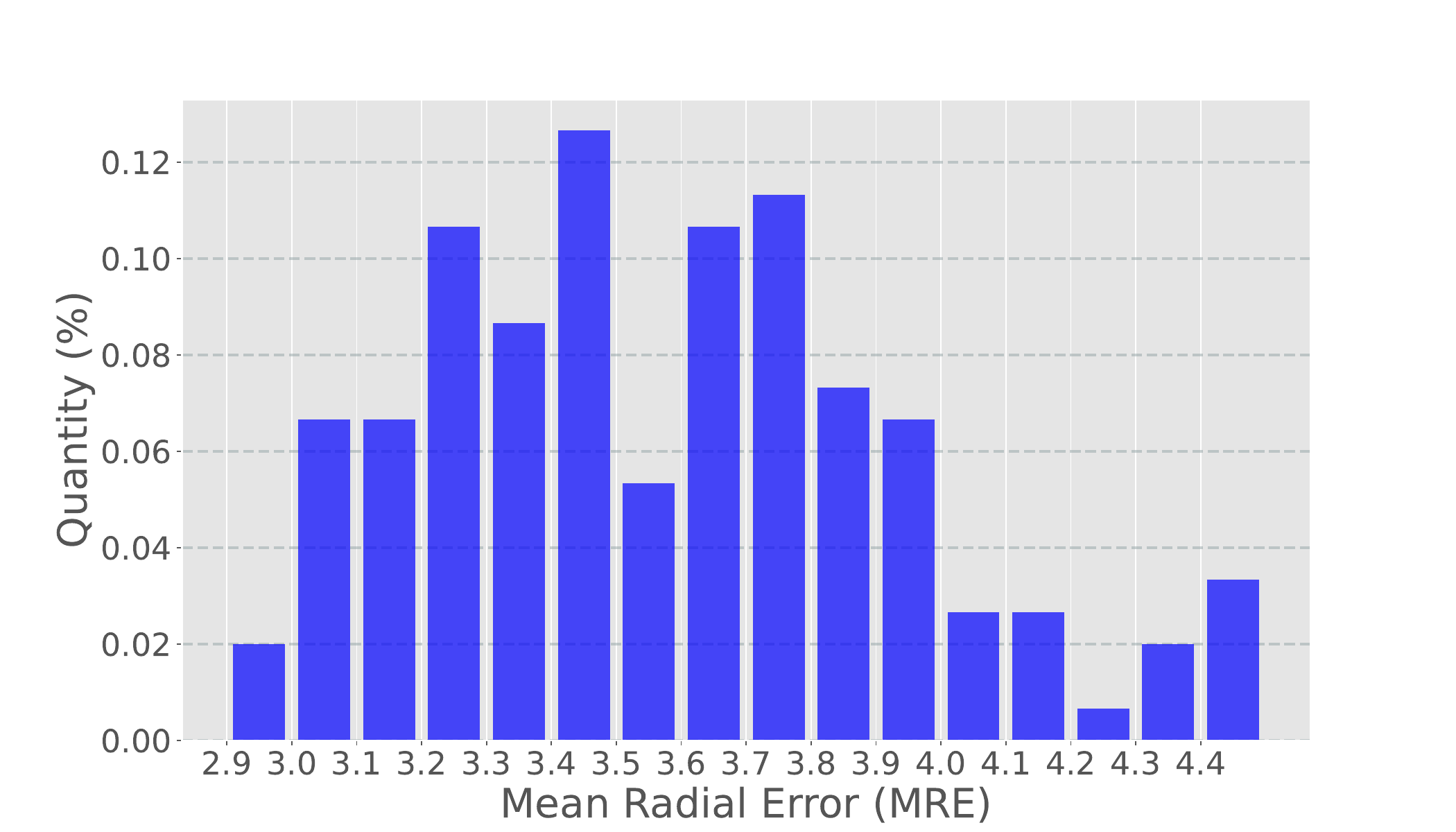}
	\caption{The distribution of the mean radial error (MRE) when choosing a different image as a template in one-shot medical landmark detection task. The x-axis refers to MRE and the y-axis refers to the percentage of MRE lying in the corresponding ranges. \textbf{Evidently, the choice of template affects the performance significantly}. }
	\label{fig:tbar_1}
\end{figure}

To alleviate this problem, many researchers~\cite{chen2021semi, laine2016temporal, DBLP:journals/corr/abs-2103-12277, tarvainen2017mean} utilize 
labeled data together with unlabeled data in a semi-supervised style to boost performance. A classic method is mean teacher~\cite{laine2016temporal,tarvainen2017mean}, 
which aggregates multiple predictions of unlabeled data by a teacher model pre-trained from labeled data. The aggregated results work as more reliable pseudo labels for unlabeled data in rest part of the method.
Another group of researchers, aiming to achieve a high performance at a low labeling cost, propose a strategy to select instances for annotation incrementally~\cite{DBLP:conf/cvpr/KimPKC21, DBLP:conf/iccv/SinhaED19, DBLP:conf/icml/TranDRC19, DBLP:conf/cvpr/Zhang0YWZH20}.
The basic idea is to first train a model with few labeled data, and then use the model to select instances from unlabeled data iteratively, which are annotated by specialists for the next round of training. 
Meanwhile, some researchers attempt to drain all potential of limited labeled data. With the power of self-training and self-supervised learning~\cite{bhalodia2020self, chen2020simple, ouyang2020self, xie2020self, yao2021label, zhou2020comparing, zhu2020rubik}, it is possible to develop a robust, few-shot model even with several labeled samples. 
For example, Yao \textit{et al.}~\cite{yao2021one} introduce a self-supervised proxy task that matches multi-layers features from images with different augmentations in the training stage, and use \textit{a single image} as the template, whose patches centered at landmarks are matched with target images to make predictions.

However, during our research following the work of \cite{yao2021one}, we observe an interesting phenomenon (see Figure~\ref{fig:tbar_1}). The template choice highly impacts the final performance. The mean radial error (MRE) of our trained model varies from \underline{2.9mm} under the ``best” template to \underline{4.5mm} under the ``worst" template. It is evident that there is a large gap lying between the best and the worst choices. Thus, a \textbf{selection question} naturally stands out: \textit{Regarding the ``gap" over samples, how to find and annotate the most ``valuable" images in order to achieve the best performance with a deep model trained under such a limited supervision?}
To the best of our knowledge, there is no ready answer to the above question. In this paper, we attempt to fill this blank. 

To answer this question, we have to address three difficulties. (1) \textbf{No supervised signal}: For a landmark detection task, there are no labels to guide our image selection --- We need to find substitutes for landmarks; 
(2) \textbf{No proper metric}: Mean radial error (MRE) is often employed as a performance metric for landmark detection, but we cannot compute MRE when no landmarks are available --- we need to find proxy for MRE;  (3) \textbf{Missing effective feature extraction}: Can we train a deep model that effectively extracts the features for template selection? 

In this paper, we propose a framework named Sample Choosing Policy (SCP) to find the most annotation-worthy images as templates. 
First, to handle the situation of no landmark label, we choose handcrafted key points as substitutes for landmarks of interest. Second, to replace the MRE, we proposed to use a similarity score between a template and the rest based on the features of such potential key points.
Third, considering landmark detection is a pixel-wise task, we apply pixel-wise self-supervised learning with all non-labeled data to build a basic feature extractor, which extracts features from each image. 
Finally, we can find out the subset of images with the highest similarities as the candidate templates to be labeled, from which a model is learned for few-shot landmark detection. With the help of SCP, we improve the MRE performance of one-shot medical landmark detection from \underline{3.595mm} (with a random template) to \underline{3.083mm} (with our selected template) in Cephalometric Xray dataset and \underline{4.114mm} (with a random template) to \underline{2.635mm} (with our selected template) in Hand Xray dataset; refer to Section~\ref{sec:experiment}.

\begin{figure}
	\centering
	\includegraphics[width=1\linewidth]{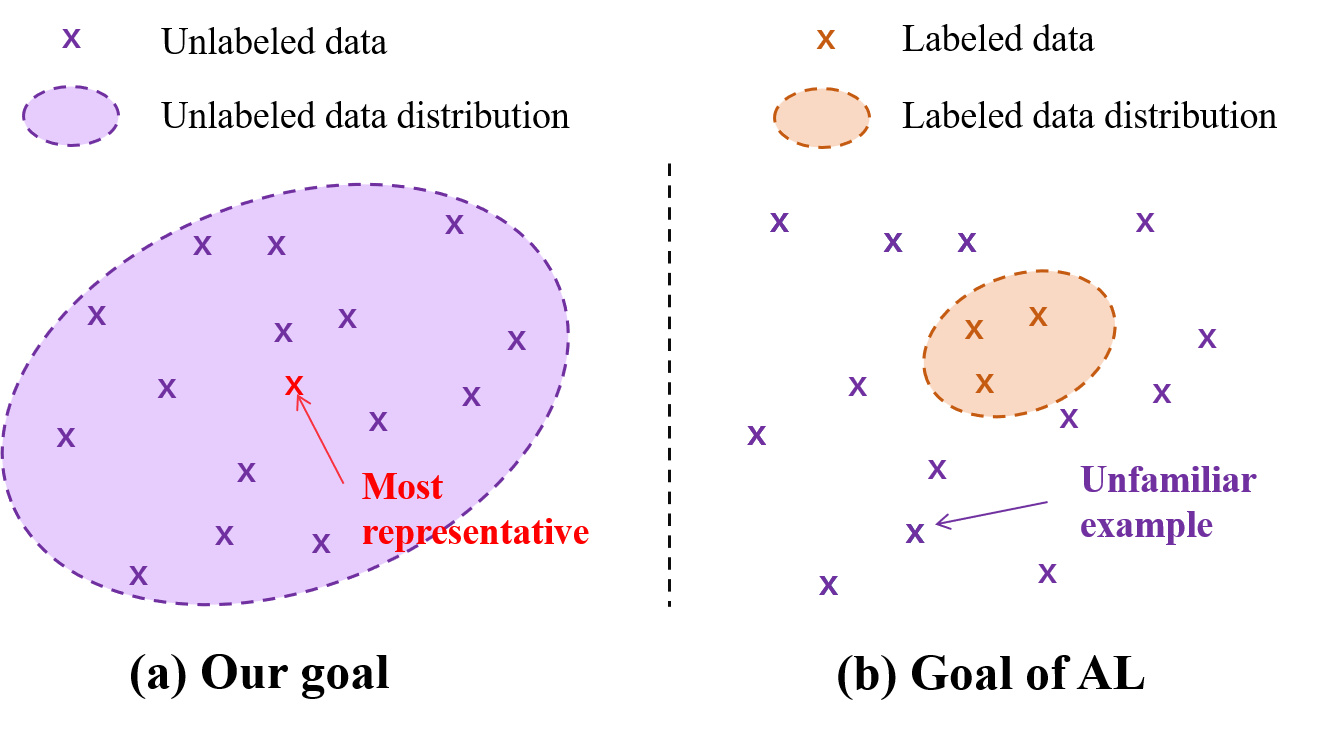}
	\caption{\textbf{Difference from active learning.} Deep models can remember and cluster the images or patterns they viewed. Instead of active learning (\textbf{AL}) tending to find the unfamiliar examples, our goal is to find the ones nearest to the center of latent space which we think more representative and important when only several images can be labeled.}
	\label{fig:dis_sample}
\end{figure}

\section{Related Work}
\subsection{Active Learning}
Active Learning queries a user interactively to label picked examples, while maximizing the performance gain of neural networks~\cite{ DBLP:conf/cvpr/GudovskiyHYT20, DBLP:conf/iclr/SenerS18, DBLP:conf/icml/SundinSSVSK19, DBLP:conf/cvpr/Zhang0YWZH20, DBLP:conf/cvpr/ZhouSZGGL17, DBLP:conf/icml/TranDRC19}. In active learning, the model is first initialized by a subset of annotated examples, and then a certain strategy is suggested to select samples to label from the unlabeled data pool. The decision of selecting a sample depends on the information gained from the model, which is measured by entropy, divergence, gradients, etc.
For example, the decision-making strategy in ~\cite{DBLP:conf/iccv/SinhaED19, DBLP:conf/icml/TranDRC19} is formed by a discriminator, assisted by a variational auto-encoder (VAE) that encodes the distribution of labeled data in a latent space. Kim et al~\cite{DBLP:conf/cvpr/KimPKC21} improve the strategy from a task perspective by updating it from a task-agnostic to task-aware style. 
More precisely, every selection in active learning aims to find hard examples in unlabeled dataset. However, the aforementioned selection question should be answered by a specific partition of samples, whose distribution is ideally as close as possible to the entire dataset, not just unfamiliar examples (Figure~\ref{fig:dis_sample}).
In addition, some active learning methods like \cite{DBLP:conf/iclr/SenerS18} also aim to collect core samples but require a labeled subset as a prerequisite. Given that, active learning is not the key to solving the selection problem.

\subsection{Self-supervised Learning}
Self-supervised learning (SSL) leverages information in data itself as the supervision, providing a solution for training from unlabeled data. Among all varieties of SSL, contrastive learning (CL) is one of the most powerful paradigms, leading to state-of-the-art performances in many vision tasks. Contrastive learning aims to pull the representations of similar samples closer, and dissimilar ones far apart. Most existing methods, like MoCo~\cite{chen2020improved, DBLP:conf/cvpr/He0WXG20}, SimCLR~\cite{DBLP:conf/icml/ChenK0H20, DBLP:conf/nips/ChenKSNH20},
BYOL~\cite{DBLP:conf/nips/GrillSATRBDPGAP20} and  BarlowTwins~\cite{DBLP:conf/icml/ZbontarJMLD21}, are designed and optimized in instance-level comparisons, benefit the trained model with more discriminative and generalized representations. But it leads to sub-optimal representations for downstream tasks requiring pixel-level prediction, e.g., segmentation, object detection, and landmark detection. Recently there are some pixel-level methods attempting to learn dense feature representations. Multi-scale pixel-wise contrastive proxy task based on InfoNCE loss~\cite{DBLP:journals/jmlr/GutmannH10, oord2018representation} is introduced in \cite{yao2021one} and achieves great performance in medical landmark detection, which is used for our feature extractor.

\subsection{Few-shot Learning}
Neural networks are often hampered when there are only a few training samples to provide supervisory signals. Few-shot learning (FSL) ~\cite{DBLP:conf/cvpr/00040ZCZ21, DBLP:conf/iclr/KwonNHL21, DBLP:conf/iclr/LiuHLJL21, DBLP:conf/iclr/RenIMZ21} is proposed to tackle this data scarcity problem, utilizing external prior knowledge to discover patterns in data. FSL has been employed successfully in tasks like segmentation~\cite{DBLP:conf/cvpr/ZhaoCL21, DBLP:conf/cvpr/NguyenT21}, detection~\cite{DBLP:conf/cvpr/SunLCYZ21, DBLP:conf/cvpr/FanMLS21}, etc.
A typical example of FSL is the pretrain-finetune paradigm. For example, Bjorn \textit{et al.}~\cite{DBLP:conf/cvpr/BrowatzkiW20} leverage image reconstruction to pretrain a model, and finetune it with a few samples for the target facial landmark task. 
For medical landmark detection, some researchers treat the few labeled samples as templates. They start with a pre-trained model from other unlabeled data and make the predictions by matching target instances and templates. Yao \textit{et al.}~\cite{yao2021one} introduce a pixel-wise multi-layer proxy task for self-supervision, 
which provides more generalized features utilized in the template-matching process. Lei \textit{et al.}~\cite{lei2021contrastive} propose another proxy task of predicting the relative offsets of two patches from an identical image. Not only that, the relative offsets also work as a metric in template matching.

\begin{figure*}[t]
	\centering
	\includegraphics[width=0.95\linewidth]{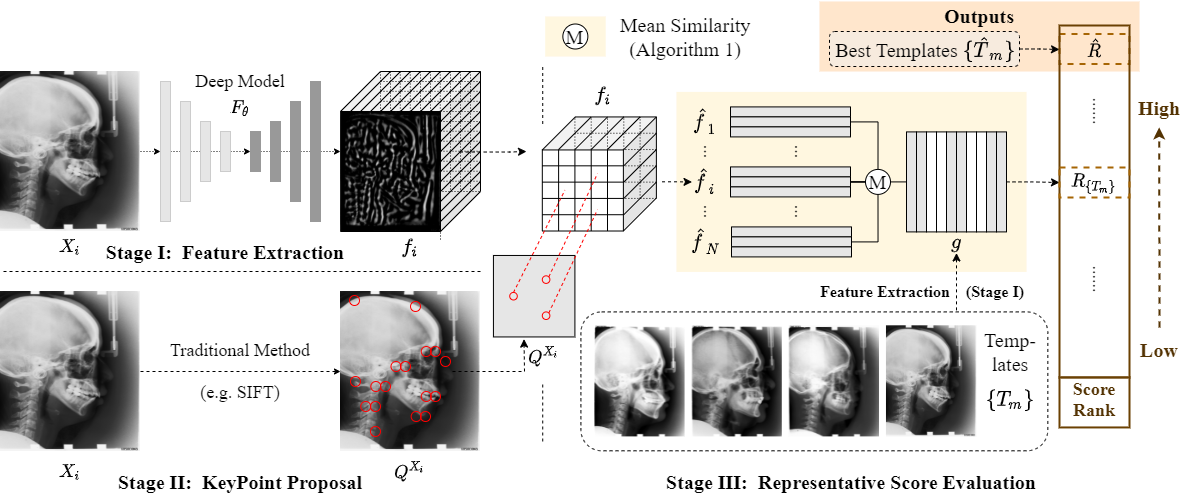}
	\caption{\textbf{Overview of Sample Choosing Policy (SCP):} SCP consists of three stages. Stage 1: Extract features $f_i$ from image $X_i$ via a self-trained deep model. Stage 2: Detect key points $Q^{X_i}$ by traditional method like SIFT from image $X_i$. Stage 3: Evaluate similarities between filtered features $\hat{f}_i$ and features $g$ of templates $\{T_m\}$, and record the average similarity $R[{\{T_m\}}]$. The combination of templates with highest similarity $\hat{R}$ are selected as $\{\hat{T}_m\}$.}
	\label{fig:overview}
\end{figure*}

\section{Method}
In this section, we introduce the proposed framework named Sample Choosing Policy (SCP) in detail. 

\subsection{Sample choosing policy} \label{sec:selection}

\paragraph{Landmark detection using template matching.} Given a template $T$, landmark detection for an image $X$ is first implemented via classical template matching. 

Denote the set of landmark by $P=\{p_1,p_2,\dots,p_L\}$. Suppose that $p_l^T \in P^T$ is the $l^{th}$ landmark point in the template $T$, its corresponding landmark $p_l^X$ in the image $X$ is found by the following \textit{searching-and-maximizing} problem:
\begin{equation}
p_l^X = \arg\max_{p} ~ s[~F_\theta \circ T(p_l^T), F_\theta \circ X(p)~]; ~p_l^T \in P^T,
\label{eq:s}
\end{equation}
where $p$ is coordinates of a pixel, $s$ is a similarity function, $F_\theta$ is a feature extractor, and $F_\theta\circ X(p)$ computes the feature map for the image $X$ and then extracts the feature vector at pixel $p$. The maximum value of the similarity function $s$ achieved by $p_l^X$ is denoted by $r_l[T\rightarrow X]$:
\begin{equation}
r_l[T\rightarrow X] = s[~F_\theta \circ T (p^T_l), F_\theta \circ X(p^X_l)~].
\label{eq:R0}
\end{equation}

The above landmark detection process considers only one template. When there are multiple templates $\{T_1,T_2,\dots,T_M\}$ indexed by $m=1:M$, template matching is implemented by
\begin{equation}
(m_l, p_l^X) = \arg\max_{(m, p)} ~ s[~F_\theta \circ T_m(p_l^T), F_\theta \circ X(p)~],
\label{eq:s_multi}
\end{equation}
which finds the best template $T_{m_l}$ for each landmark $l$ as well as the matched landmark location $p_l^X$. By the same token,
the maximum value of the similarity function $s$ achieved by $(m_l, p_l^X)$ is denoted by $\hat{r}_l[\{T_m\}\rightarrow X]$:
\begin{equation}
\hat{r}_l[\{T_m\}\rightarrow X] = s[~F_\theta \circ T_{m_l} (p^T_l), F_\theta \circ X(p^X_l)~].
\label{eq:R}
\end{equation}

For the choice of \textit{similarity function}, we utilize the commonly used cosine similarity function:
\begin{equation}
s[~v^T,v^X~] = CosSim(v^T,v^X) = \frac{\langle v^T \cdot v^X \rangle}{||v^T||_2 \cdot ||v^X||_2},
\label{eq:cos}
\end{equation}
where $v$ is a feature vector. 

\paragraph{Template selection.} 

To single out the best $M$ templates among a set of images $\Omega=\{X_1,X_2,\ldots,X_N\}$, we aim to seek the set of templates $\{T_m\}$ that contains ``the most similar" landmark information for all landmarks and with respect to all images:
\begin{equation}
\{{\hat T}_m\} = \arg \max_{ \{T_m\}  \subset \Omega}~  \frac{1}{N} \sum_{n} \frac{1}{L} \sum_l {\hat r}_l [ \{T_m\} \rightarrow X_n ]. \label{eq:T}
\end{equation}
The above optimization is \textit{combinatorial in nature} as there are $\binom{N}{M}$ possible combinations, which are nearly impossible to exhaust in practice except for very small $M$. Therefore, we randomly sample a large number (say $10,000$) of combinations and pick the maximizing combination as an approximate solution.

\paragraph{Substituting landmarks with key points.} \label{subsec:filter} To implement template selection per Eq. (\ref{eq:T}), the knowledge of landmarks is assumed. However, even such knowledge is nonexistent before template selection. Therefore, we proposed to utilize potential key points to substitute landmarks. In particular, we utilize the classical multi-scale detector, SIFT, to find key points, where landmarks are likely to co-locate.

For each image $X \in \Omega$, we apply SIFT to get its corresponding $K$ key points $Q^X = \{q^X_1, q^X_2, \dots, q^X_K\}$ with the highest responses.
Further, the SIFT key points for different images are not in correspondence, directly applying Eq. (\ref{eq:s_multi}) is not possible. To address such an issue, we perform the template matching in a \textit{reverse order}, that is, for an image $X$ with its key points $Q^X$, we perform the following for each key point $q_k^X$:
\begin{equation}
(m_k, q_k^T) = \arg\max_{(m, q)} ~ s[~F_\theta \circ T_m(q), F_\theta \circ X(q_k^X)~],
\label{eq:s_multi2}
\end{equation}
and record the achieved maximum as 
\begin{equation}
\hat{r}_k[X \rightarrow \{T_m\} ] = s[~F_\theta \circ T_{m_k} (q^T_k), F_\theta \circ X(q^X_k)~].
\label{eq:R2}
\end{equation}
Finally, we define the average similarity $R$ as the representative score of $\{T_m\}$:
\begin{equation}
R[{\{T_m\}}] = \frac{1}{N} \sum_{n} \frac{1}{K} \sum_k {\hat r}_k [ X_n \rightarrow  \{T_m\}  ]. \label{eq:T2}
\end{equation}
and Eq. (\ref{eq:T}) is accordingly adapted:
\begin{equation}
\{{\hat T}_m\} = \arg \max_{ \{T_m\}  \subset \Omega}~R[\{T_m\}]
\end{equation}

\paragraph{Construction of feature extractor using contrastive learning.}
\label{subsec:extractor}
To answer the question that what kind of deep model can support us to make selection, we start our analysis of Eq.~(\ref{eq:s}), which maximizes the similarities between the same landmarks from different images. 

Without landmark labeling, we resort to contrastive learning, which is proven to be a reliable tool to learn a basic model without using any label information. Here, instead of instance-level self-supervised learning for visual recognition \cite{DBLP:conf/nips/GrillSATRBDPGAP20,DBLP:conf/icml/ZbontarJMLD21}, we adapt it for achieving the goal of (\ref{eq:s}). We do so by narrowing the distance between different views of an identical patch and extend the distance between different patches. For example, InfoNCE loss~\cite{oord2018representation} is widely used and applied in our training for feature extractor,
\begin{equation}
\begin{split}
\mathcal{L}_{\text{InfoNCE }} &= -\mathbb{E} \left[\log \frac{\exp(\alpha)}{\exp(\alpha)+\sum \exp(\alpha')}\right];\\
\alpha &= s[~F_\theta \circ X (p), F_\theta \circ X_{aug} (p)~];\\
\alpha' &= s[~F_\theta \circ X(p), F_\theta \circ X(q)~],
\end{split}
\end{equation}
where $X_{aug}$ is a different version of $X$ by augmentation, and $p$ and $q$ are two different key points. 

We follow \cite{yao2021one,yao2020miss} to construct a deep model trained via multi-layer pixel-wise contrastive loss function as our feature extractor. According to \cite{yao2021one}, we use VGG~\cite{DBLP:journals/corr/SimonyanZ14a} as the backbone, followed with 5 blocks to reduce the dimension. This model is trained with a pixel-wise matching proxy task for over 500 epochs.

\paragraph{The overall template selection pipeline.} \label{subsec:overall}
Based on the above discussion, our pipeline is summarized as in Figure~\ref{fig:overview}, assuming the availability of the feature extractor $F_\theta$, which is learned using the self-supervised pixel-wise matching task. First, we extract features from all images $\Omega$ and candidate templates $\{T_m\}$ for the following operations. Second, we extract key points $Q^X$ with the help of traditional key point detector SIFT. Third, we pair each image $X_n$ and the template group $\{T_m\}$ to obtain the similarity $r_n$ between $X_n$ and $\{T_m\}$ (Eq.~(\ref{eq:R2})). The mean similarity $R_{\{T_m\}}$ of all pairs of $X_n$ and $\{T_m\}$ indicate the ``representativeness" of $\{T_m\}$ to the whole dataset, and the best group of templates $\{\hat{T}_m\}$ achieving the maximum of mean similarity are our final selection (Eq.~(\ref{eq:T2})).

\subsection{Refined landmark detection}
\label{subsec:refine}
While landmark detection is implemented as template matching in Section \ref{sec:selection}, its detection performance is still limited as its feature detector is geared for all pixels not specifically for the landmarks. We further follow \cite{yao2021one} to improve the detection of landmarks via semi-supervised learning. Another landmark detection deep model with a heatmap predictor and two offset predictors (offsets in x- and y-axis) is built for distilling with pseudo landmark labels predicted by the previous template matching model. The performance of the distilled model is geared toward landmarks of interest and is better than the previous model.

\begin{figure}
	\includegraphics[width=0.98\linewidth]{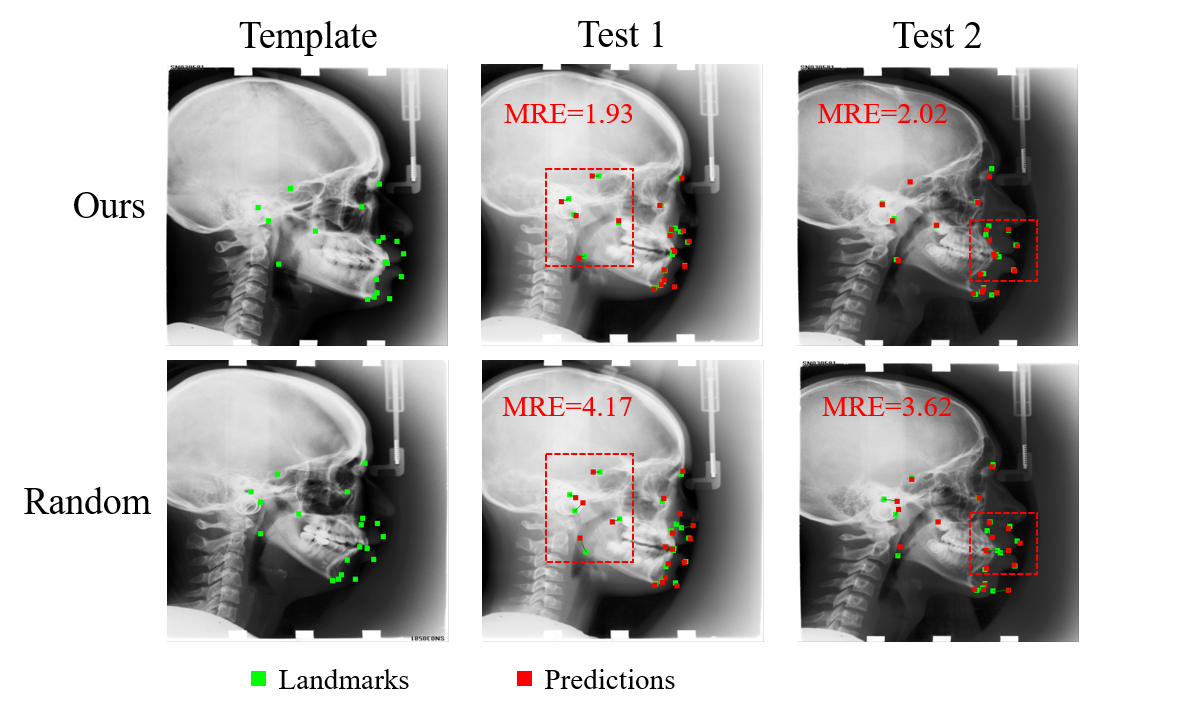}
	\caption{\textbf{Visual Comparison of templates from our policy and random selection.} Column ``Template/Test 1/Test 2" refers to the templates and two test images. The row ``Ours" and ``Random" refers to the template selected by our method and random selection, respectively.
		As shown in \textcolor{red}{red} dashed boxes, our template outperforms the random selected template in visualization. }
	\label{fig:demo}
\end{figure}

\section{Experiments}
\label{sec:experiment}
\subsection{Datasets}
\bheading{Cephalometric Xray:}
It is a widely-used public dataset for cephalometric landmark detection, containing 400 radiographs, and is provided in IEEE ISBI 2015 Challenge~\cite{link_dataset,wang2016benchmark}. There are 19 landmarks of anatomical significance labeled by 2 expert doctors in each radiograph. The averaged version of annotations by two doctors is set as the ground truth. The image size is $1935 \times 2400$ and the pixel spacing is 0.1mm. The dataset is split into 150 and 250 for training and testing respectively, referring to the official division.

\bheading{Hand Xray:}
It is also a public dataset including 909 X-ray images of hands. The setting of this dataset follows ~\cite{DBLP:journals/mia/PayerSBU19}. The first 609 images are used for training and the rest for testing. The image size varies among a small range, so all images are resized to 384$\times$384.

\bheading{WFLW:} 
This dataset is from \cite{DBLP:conf/cvpr/Wu0YWC018} containing 10,000 faces with 7500 and 2500 in training and test sets, respectively. All images are collected from the WIDER FACE dataset \cite{DBLP:conf/iccv/YangLOLW17} and manually labeled with 98 landmarks. The dataset contains different test subsets where the image appearances vary due to variations in pose, expression, and/or illumination or the presence of blur, occlusion, and/or make-up.

\subsection{Settings}
\bheading{Metrics:} Following the official challenge~\cite{link_dataset,wang2016benchmark}, mean radial error (MRE) and successful detection rate (SDR) in four radii (2mm, 2.5mm, 3mm, and 4mm) are applied, based on the Euclidean distance between prediction and ground truth. In addition, similarity via Eq. (\ref{eq:T2}) is demonstrated for comparison. For the WFLW dataset, the performance of facial landmark detection is reported by normalized mean error (NME), failure rate at 10\% NME (FR@10), and area-under-the-curve (AUC) of the Cumulative Error Distribution (CED) curve.

\bheading{Implementation details:}
All of our models are implemented in PyTorch, accelerated by an NVIDIA RTX GPU.
Following~\cite{yao2020miss, yao2021one}, the feature extractor is optimized by Adam optimizer for 3500 epochs for self-supervised training. The learning rate is initialized with 0.001, and decayed by half every 500 epochs. It takes ~6 hours to converge with batch size set to 8. 

\subsection{The performance of SCP}

\begin{table*}[t]
	\centering
	\caption{Comparison of different numbers of templates on the ISBI 2015 Challenge~\cite{wang2016benchmark} and Hand Xray~\cite{yao2021one} testsets. The landmarks are detected based template matching.}
	\begin{tabular}{lrrrrrrr}
		\hline
		\multirow{2}{*}{Dataset} & \multirow{2}{*}{M} & \multicolumn{2}{c}{Similarity ($\uparrow$)} & \multicolumn{4}{c}{MRE ($\downarrow$) (mm)} \\
		&  & ours & random & ours & random & best & worst \\
		\hline
		Cephalometric & 1   & 0.554 &0.519$\pm$0.017&   3.083 &3.595$\pm$0.381&2.863&4.952 \\
		& 2   & 0.638 &0.604$\pm$0.012&   2.840 &3.177$\pm$0.227&2.661&4.090 \\
		& 3   & 0.676 &0.646$\pm$0.009&   2.742 &3.019$\pm$0.187&2.573&3.875 \\
		& 4   & 0.699 &0.672$\pm$0.008&   2.677 &2.912$\pm$0.164&2.512&3.674 \\
		& 5   & 0.709 &0.690$\pm$0.007&   2.571 &2.850$\pm$0.141&2.486&3.458 \\
		
		\hline
		Hand Xray & 1  & 0.363 &  0.288$\pm$0.052  &2.635 &  4.114$\pm$3.566 & 2.468 & 43.81 \\
		& 5  & 0.532 &  0.482$\pm$0.021  &2.891 &  3.188$\pm$0.493 & 2.377 & 6.268  \\
		& 10 & 0.601 &  0.547$\pm$0.013  &2.769 &  3.070$\pm$0.346 & 2.289 & 4.808  \\
		& 15 & 0.615 &  0.581$\pm$0.011  &2.664 &  2.989$\pm$0.311 & 2.259 & 4.245  \\ 
		& 50 & 0.690 &  0.673$\pm$0.006  &2.524 &  2.740$\pm$0.219 & 2.240 & 3.775  \\ 
		\hline
	\end{tabular}\label{table:ceph_hand}
\end{table*}

\begin{table*}
	\centering
	\caption{Comparison of the state-of-the-art landmark detection approaches and our refined approach on the ISBI 2015 Challenge~\cite{wang2016benchmark} testset. } 
	\begin{threeparttable}
		\begin{tabular}{lrrrrrr}
			\hline
			\multirow{2}{*}{Model} & \multirow{2}{*}{\tabincell{l}{Labeled \\ images}} & 
			\multirow{2}{*}{\tabincell{c}{MRE ($\downarrow$) \\ (mm)}} &  \multicolumn{4}{c}{SDR ($\uparrow$) (\%)} \\ \cline{4-7}
			&  &  & 2mm & 2.5mm & 3mm & 4mm \\ \hline
			Ibragimov \textit{et al.}~\cite{ibragimov2015computerized}*  & 150 & - & 68.13 & 74.63 & 79.77 & 86.87\\
			Lindner \textit{et al.}~\cite{lindner2015fully}* & 150 & 1.77 & 70.65 & 76.93 & 82.17 & 89.85\\
			Urschler \textit{et al.}~\cite{ref_urschler}* & 150 & - & 70.21 & 76.95 & 82.08 & 89.01\\
			Payer \textit{et al.}~\cite{ref_scn}* & 150 & - & \textbf{73.33} & \textbf{78.76} & \textbf{83.24} & \textbf{89.75}\\
			\hline
			Payer \textit{et al.}~\cite{ref_scn}$^{\dagger}$ & 25 & 2.54 & \textbf{66.12} & \textbf{73.27} & \textbf{79.82} & 86.82\\
			Yao \textit{et al.}~\cite{yao2021one}$^{\dagger}$ & 25 & 2.17 & 55.83 & 66.65 & 76.88 & 89.13\\
			SCP (Ours w/ refinement) & 25 & \textbf{2.06} & 58.00 & 68.52 & 79.13 & \textbf{90.40}\\
			\hline
			Payer \textit{et al.}~\cite{ref_scn}$^{\dagger}$ & 10 & 6.52 & 49.49 & 57.91 & 65.87 & 75.07\\
			Yao \textit{et al.}~\cite{yao2021one}$^{\dagger}$   & 10 & 2.30 & 53.11 & 63.58 & 74.34 & 87.17\\
			SCP (Ours w/ refinement)  & 10 & \textbf{2.24} & \textbf{55.41} & \textbf{64.67} & \textbf{75.26} & \textbf{87.38}\\
			\hline
			Payer \textit{et al.}~\cite{ref_scn}$^{\dagger}$ & 5 & 12.34 & 27.35 & 32.94 & 38.48 & 45.28\\
			Yao \textit{et al.}~\cite{yao2021one}$^{\dagger}$ & 5 & 2.44 & 49.92	& 60.83	& 71.76	& 85.13 \\
			SCP (Ours w/ refinement)  & 5  & \textbf{2.33}  & \textbf{52.25} & \textbf{63.05} & \textbf{73.95} & \textbf{86.27} \\
			\hline
			Yao \textit{et al.}~\cite{yao2021one}$^{\dagger}$ & 1 & 2.90 & 37.16 & 48.04 & 60.02 & 77.72\\
			SCP (Ours w/ refinement)  & 1 & \textbf{2.74} & \textbf{43.79} & \textbf{53.05} & \textbf{64.12} & \textbf{79.05}\\
			\hline
		\end{tabular}
		
		\begin{tablenotes}
			\footnotesize
			\item[*] copied from their original papers.
			\item[$\dagger$] re-implemented with limited labeled images.
		\end{tablenotes}
	\end{threeparttable}
	\label{Table:Main}
\end{table*}

\bheading{Few-shot medical landmark detection:} Firstly, experiments are conducted on different numbers of templates for Cephalometric dataset. For Table~\ref{table:ceph_hand}, $M$ denotes the number of templates used in experiment. The columns ``ours" refer to the results achieved by the proposed method, while ``random" means to the average results of multiple rounds of training (we use 1,000 trials), with standard variations. The column "best" and "worst" refer to the best or worst results in multiple tries, respectively.

As shown in Table~\ref{table:ceph_hand}, our approach cannot find the best templates, instead, it finds a fairly good choice well above the average performance. When there is only one template, the best template achieves $2.863$mm in MRE. Although the template we choose achieves an MRE of only $3.083$mm, it is much better than the average MRE $3.595$mm. Another fact revealed by Table~\ref{table:ceph_hand} is the ``diminishing return" phenomenon: As the number of templates increases, the difference of maximum similarities and performances between choices tends to be smaller. In our experiment, we find that when $M$ achieves 75 (half of the dataset) or more, our template selection policy has little effect because of sufficient modeling of the appearance variance in medical images.

The predictions by one template from our method and random selection are visualized in Figure~\ref{fig:demo}. Our predictions around ears and nose locate more closer to the ground-truth landmarks than those by random template, which has consistent performance in MRE, quantitatively.

Besides, another group of experiments are conducted on Hand Xray dataset. The proposed SCP is applied on Hand Xray dataset to obtain the suggested $M$ templates. Following settings in \cite{yao2021one}, evaluation model is built. Results are listed in Table~\ref{table:ceph_hand}, showing reliable improvements (e.g., MRE reduced by $35.5\%$ ($4.114$mm to $2.653$mm)) . As reported above, MRE results of 5/10/15 templates perform a bit worse than that of 1 template (e.g., the MRE is $2.891$mm for 5 templates, but $2.653$mm for 1 template). The dataset size of Hand Xray (609 images) is much bigger than that of Cephalometric (150 images). We speculate that the diversity of Hand Xray dataset could not be well "represented" by such small group of templates. The smaller the number of templates is, the more randomness it underlies. So it makes sense the results tend to be stable when the number of templates increases.

\bheading{Refined landmark detection results:} To compare with supervised landmark detection methods, we follow \cite{yao2021one} and apply semi-supervised refinement based on the self-supervised results to achieve better performance (refer to Section~\ref{subsec:refine}). Our methods and \cite{yao2021one} are trained with labeled and unlabeled data, while \cite{ibragimov2015computerized,lindner2015fully,ref_scn,ref_urschler} are trained with labeled images . As in Table \ref{Table:Main} our method outperforms others when labels are limited (e.g., the MRE is $2.33$mm for SCP with 5 labeled images, $2.44$mm for \cite{yao2021one}, and $12.34$mm for \cite{ref_scn}), and achieves detection results comparable to the supervised methods that are trained based on even more labeled images (e.g., the MRE for \cite{ref_scn} with 25 labeled images is $2.54$mm).

\bheading{Facial landmark detection:}
We attempt to apply our templates selection method for \cite{DBLP:conf/cvpr/BrowatzkiW20} to improve the performance on facial landmark detection. The pre-trained model from the first stage of \cite{DBLP:conf/cvpr/BrowatzkiW20} is integrated into the proposed framework, which suggest the $M$ instances with the highest mean similarities ($M$=50 in our experiment). In order to eliminate unstable factors, models with the first $M$ instances or instances from our policy are both trained for 5 times.
The evaluation is conducted on full and make-up test split of WFLW dataset. The results are listed in Table~\ref{table:wflw} where our method significantly improves NME from $9.131$ to $8.998$ in full split and from $8.724$ to $8.578$ in make-up split.

\begin{table}[t]
	\centering
	\caption{Improvements on WFLW dataset based on \cite{DBLP:conf/cvpr/Wu0YWC018} with 50 labeled images.}
	\begin{threeparttable}
		\begin{tabular}{lrrr}
			\hline
			\multicolumn{4}{c}{WFLW dataset}  \\
			\hline
			Full   & NME($\downarrow$)     & FR@10($\downarrow$)   & AUC($\uparrow$)  \\
			\hline
			$\text{Original}^{\dagger}$ & 9.131$\pm$0.316& 27.506$\pm$1.785 & 0.278$\pm$0.021 \\
			Ours   & \textbf{8.998$\pm$0.129}& \textbf{27.475$\pm$0.840} & \textbf{0.285$\pm$0.002} \\
			\hline
			
			Make-up & NME($\downarrow$)     & FR@10($\downarrow$)    & AUC($\uparrow$)    \\
			\hline
			$\text{Original}^{\dagger}$ & 8.724$\pm$0.054& 25.79$\pm$1.689 & 0.283$\pm$0.003 \\
			Ours& \textbf{8.578$\pm$0.188} & \textbf{25.79$\pm$0.989} & \textbf{0.284$\pm$0.003}  \\
			\hline
		\end{tabular}
		\begin{tablenotes}
			\footnotesize
			\item[$\dagger$] re-implemented from \cite{DBLP:conf/cvpr/BrowatzkiW20} 
		\end{tablenotes}
	\end{threeparttable}
	\label{table:wflw}
\end{table}

\begin{table*}[h]
	\centering
	\caption{Comparison for different SSL methods and different traditional key point detectors on Cephalometric dataset.} 
	\begin{tabular}{c|ccc|cccc}
		\hline
		\multirow{3}{*}{M} & \multicolumn{7}{c}{MRE ($\downarrow$) (mm)}\\
		\cline{2-8}
		& \multicolumn{3}{c}{SSL methods}  & \multicolumn{4}{|c}{Key Point Detectors}  \\
		& Ours & BYOL-m~\cite{DBLP:conf/nips/GrillSATRBDPGAP20} & BYOL~\cite{DBLP:conf/nips/GrillSATRBDPGAP20}  & SIFT &  SURF &  ORB &  Random \\
		\hline
		1  & \textbf{3.083} & 3.366 & 3.362 & \textbf{3.080} & 3.208 & 3.668 & 3.181  \\
		2  & \textbf{2.846} & 3.086 & 3.085 & \textbf{2.846} & 2.901 & 3.296 & 3.219  \\
		3  & \textbf{2.743} & 3.008 & 2.877 & \textbf{2.743} & 2.787 & 2.896 & 2.860  \\
		4  & \textbf{2.673} & 2.894 & 2.761 & \textbf{2.673} & 2.621 & 2.621 & 2.887  \\
		5  & \textbf{2.571} & 2.762 & 2.739 & \textbf{2.571} & 2.571 & 2.677 & 2.835  \\
		\hline
	\end{tabular}
	\label{table:abla}
\end{table*}

\section{Ablation Study}

\begin{table}[t]
	\centering
	\caption{Ablation study of different number of SIFT key points and searches for selecting best templates on Cephalometric dataset.}
	\begin{tabular}{lrrr}
		\hline
		Ablation & Num & Similarity ($\uparrow$) & MRE ($\downarrow$) (mm)\\ 
		\hline
		Keypoints & 50   & 0.7128 & 2.670 \\ 
		& 75   & 0.7099 & 2.623  \\ 
		& 100  & \textbf{0.7117} & \textbf{2.571} \\ 
		& 150  & 0.7105 & 2.599  \\ 
		\hline
		Searches &1000 & 0.7080  & 2.711 \\ 
		&10000        & 0.7105 & 2.703  \\ 
		&50000        & 0.7108 & 2.577 \\ 
		&100000       & \textbf{0.7117} & \textbf{2.571} \\
		\hline
	\end{tabular}
	\label{table:abla2}
\end{table}

\subsection{Key Point Proposal}
\begin{figure}[t]
	\centering
	\includegraphics[width=1\linewidth]{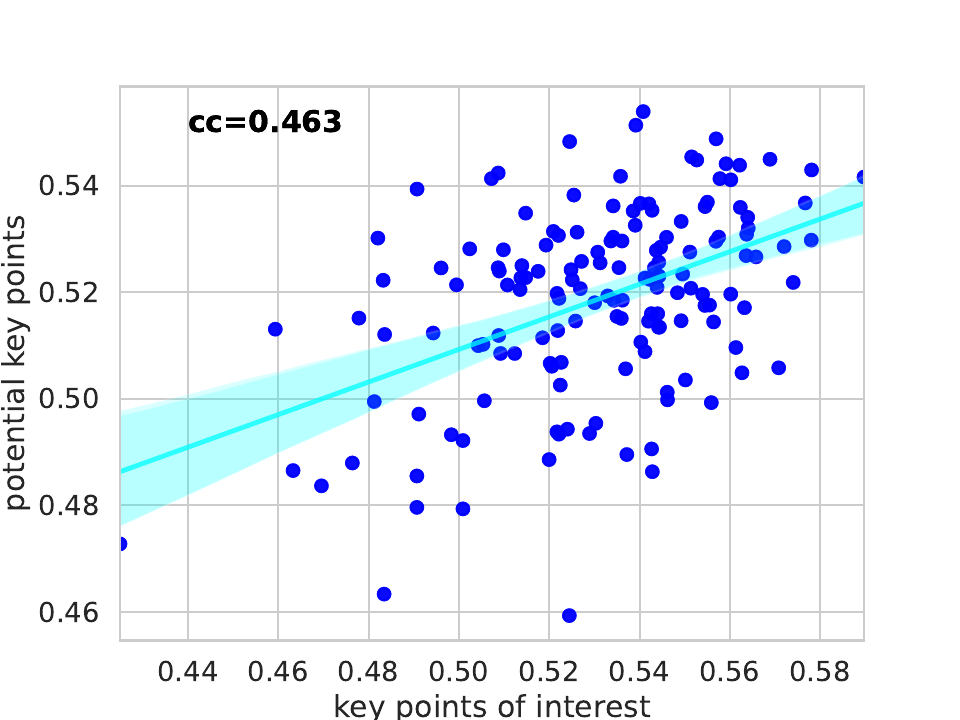}
	\caption{\textbf{Similarities of potential key points vs. landmarks.} The correlation coefficient (CC) of potential key points and landmarks is 0.462, thus we think it is feasible to replace landmarks with potential key points when estimating similarities.}
	\label{fig:sim2}
\end{figure}

\textit{Q: How good is the use of SIFT key points as substitutes for landmarks?} Figure~\ref{fig:sim2} demonstrate the relationship between landmarks and potential key points from handcraft methods in feature level (Eq. (\ref{eq:T2})). 
The similarities calculated by potential key points are positively correlated with those by landmarks to a large extent. Thus, we pick the potential key points as the substitutes.
It should be noted that excessive key points will not improve performance but increase the computational cost (e.g. $2.571$mm for 100 key points and $2.599$mm for 150 key points in Table~\ref{table:abla2}), but sparse key points will lead to performance degradation (e.g. MRE increase from $2.571$mm to $2.670$mm as the number of key points decrease from 100 to 50.) 

\textit{Q: Do other kinds of handcrafted key points also work in our framework?} In the proposed method, we use SIFT points as the potential key points to avoid estimating features of all pixels. 
To compare with other key point detection methods, we adopt SIFT, SURF, ORB, and random selection as our key point selectors. For SIFT, SURF, and ORB, we detect the key points and then filter out some very close points. For random selection, we randomly select 100 points as the key points.
The results are listed on Table~\ref{table:abla}, where we find the performance of SURF is close to SIFT, and both greatly exceed the random filtering (e.g., SIFT/SURF are both $2.571$mm for 5 templates, and greater than $2.835$mm of random points). The locations of ORB key points concentrate in several area and performs less robust than SIFT and SURF. 

\begin{table}[]
	\centering
	\caption{Comparison with different sampling methods on Cephalometric dataset.}
	\begin{tabular}{lrr}
		\hline
		Method         & Similarity($\uparrow$) & MRE($\downarrow$) (mm)   \\
		\hline
		Ours              & \textbf{0.709}  & \textbf{2.571} \\
		VAAL~\cite{DBLP:conf/iccv/SinhaED19}               & 0.691$\pm$0.004 & 2.851$\pm$0.077\\
		Loss(MSE)         & 0.695$\pm$0.008 & 2.766$\pm$0.093  \\
		Uncertainty       & 0.694$\pm$0.005 & 2.876$\pm$0.074 \\
		Entropy           & 0.692$\pm$0.006 & 2.847$\pm$0.094 \\
		\hline
	\end{tabular}
	\label{table:al}
\end{table}

\begin{figure}[t]
	\centering
	\includegraphics[width=1\linewidth]{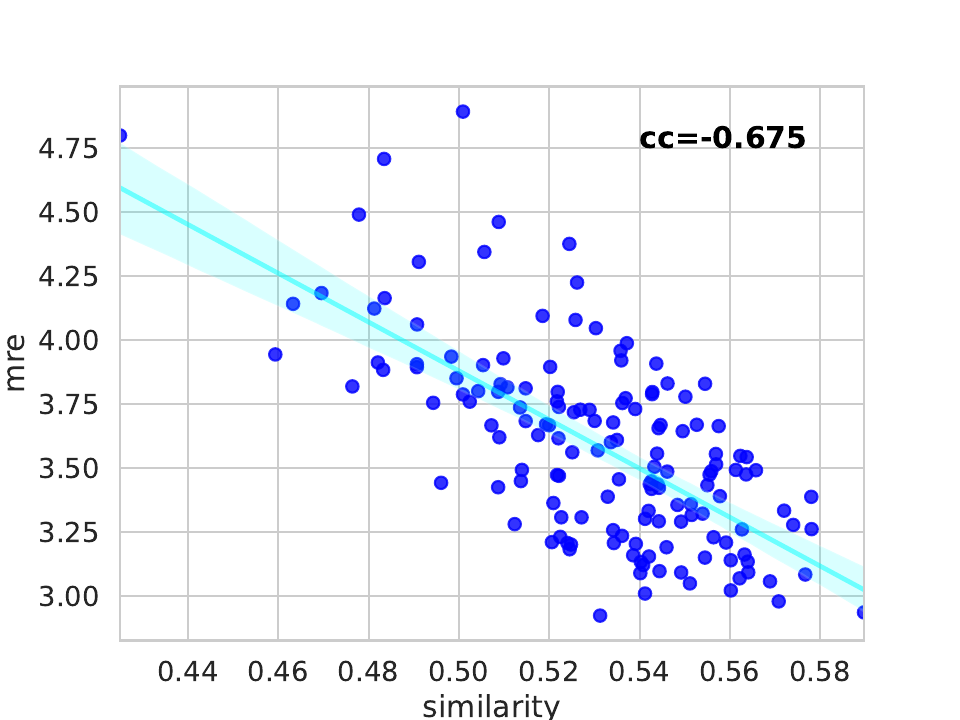}
	\caption{\textbf{Mean Similarity vs. Mean Radial Error (MRE).} Similarities are estimated by calculating the average similarities of all landmarks. The correlation coefficient (CC) of the two indicators is -0.675, indicating a strong, negative correlation.}
	\label{fig:max_sim_1}
\end{figure}

\subsection{Selection Policy}
\textit{Q: How good is the proposed similarity correlated with MRE?} The metric of MRE is widely used for evaluating the landmark detection performance. Therefore, it is of interest to know how the proposed similarity is correlated with MRE. Figure~\ref{fig:max_sim_1} presents such a plot, which nicely shows a strong (negative) correlation coefficient of $-0.675$. This enables us to utilize the proposed similarity function as a proxy of MRE, which is more convenient for us to estimate when there is no label.

\textit{Q: What about other selection methods?} We also try some other methods based on ideas from active learning. VAAL~\cite{DBLP:conf/iccv/SinhaED19} is a typical active learning framework which we re-implement and start with random 5 instances. Next, we obtain $N$($N=5$) instances suggested by VAAL to be templates.
Uncertainty and Entropy are also common tools for active learning~\cite{DBLP:journals/corr/abs-2009-00236}.
We re-implement the framework by connecting one encoder with two classifiers, and train by classifying all instances and enlarging the difference of outputs from two classifiers. 
Finally, we estimate the entropy (of one-hot vectors), uncertainty (difference of two classifiers) and loss as metrics to suggest templates to label.
As shown in Table~\ref{table:al}, VAAL needs labeled data to initialize, and performs badly in only one iteration. 
MSE works while uncertainty and entropy are not, where the probable reason is that patterns in medical images appear simple for both classifiers which give similar predictions and cause very low uncertainty for all images and low entropy to distinguish instances clearly. 

\textit{Q: How does the number of combinations affect the final performance?} As the number of templates increase, it becomes much harder to iterate all combinations. Therefore, we randomly choose the templates in limited iterations to find the best combination. We conduct an experiment about the number of random selections which is shown in Table~\ref{table:abla2}. As the number of random selections increases, we have more chances to find the combination with a larger similarity, which yields better landmark detection performance.

\subsection{Self-supervised Method}

\textit{Q: Can self-supervised methods affect our performance?} In the proposed method, we leverage the proxy task in \cite{yao2021one} to pre-train our feature extractor. In addition, we implement other self-supervised methods, including BYOL~\cite{DBLP:conf/nips/GrillSATRBDPGAP20} in pixel level, and BYOL with multi-layer training (BYOL-m). 
From Table~\ref{table:abla} we discover the strong correlation between the performance of self-supervised models and the quality of selection results and our model with multi-layer InfoNCE outperforms others in feature extraction (e.g. 3.083mm for ours and 3.366/3.263mm for others) due to more fitting the landmark detection task. 

\section{Conclusion and Future work}

We propose Sample Choice Policy (SCP) for few-shot medical landmark detection task, a novel framework for screening out the representative instances to reduce labor on annotation and achieve high performance simultaneously. SCP learns to map the consistent anatomical information into feature spaces by solving a self-supervised proxy task and extracting representative patches from all images in the first stage. Simultaneously, SCP leverages traditional key point detector to pick out valuable patches with a large local variation or steep edges in images as the representatives. Finally, SCP estimates the relevance between images by averaging the similarities between their representative patches, and selects the images with high average relevance with all other images. Our extensive experiments show that SCP outperforms the conventional policies of template selection and, after refinement, achieves state-of-the-art performances for few-shot medical landmark detection. In the future, we plan to further improve the efficiency and accuracy of finding out the best templates by designing deep models that extract more representative features and to explore the idea of template selection for other applications such as image segmentation.

\newpage
{\small
\bibliographystyle{ieee_fullname}
\bibliography{Main}
}

\end{document}